\documentclass[aps,prd,twocolumn,nofootinbib,floatfix]{revtex4-1}
\usepackage{setspace}
\usepackage{amsthm}
\theoremstyle{definition}

\usepackage{bbold}
\usepackage{graphicx}
\usepackage{dcolumn}
\usepackage{multirow}
\usepackage{subfigure}
\usepackage{times,mathptm}
\usepackage{float}
\usepackage{color}
\usepackage{amsmath,amsfonts}
\usepackage{mathptmx}
\usepackage{mathrsfs}
\usepackage{bbm}
\usepackage{bm}
\usepackage{xfrac}

\newcommand{\beq}{\begin{equation}}
\newcommand{\eeq}{\end{equation}} 
\newcommand{\bea}{\begin{eqnarray}}
\newcommand{\eea}{\end{eqnarray}}

\newcommand{\E}{\mathcal{E}_0}

\renewcommand{\d}{\delta}

\renewcommand{\l}{\lambda}

\renewcommand{\b}{\beta}

\newcommand{\tr}{\text{Tr}}

\newcommand{\vx}{{\vec{x}}}
\newcommand{\vy}{{\vec{y}}}
\newcommand{\vz}{\vec{z}}

\newcommand{\m}{\mu}
\newcommand{\pbar}{\overline{\psi}}

\newcommand{\g}{\gamma}

\newcommand{\s}{\sigma}

\renewcommand{\th}{\theta}

\newcommand{\oh}{\frac{1}{2}}

\newcommand{\dg}{\dagger}
\newcommand{\non}{\nonumber}
\renewcommand{\t}{\tau}
\newcommand{\rf}[1]{(\ref{#1})}
\newcommand{\ra}{\rightarrow}
\newcommand{\pa}{\partial}
\renewcommand{\vec}[1]{\bm #1}

\usepackage{ulem}

\begin{document}

\title{Does the Z boson have a lighter cousin?} 

\bigskip
\bigskip

\author{Jeff Greensite}
\affiliation{Physics and Astronomy Department \\ San Francisco State
University  \\ San Francisco, CA~94132, USA}
\bigskip
\date{\today}
\vspace{60pt}
\begin{abstract}

\singlespacing
 
        In the quenched electroweak theory on the lattice I construct a set of physical states which overlap the physical photon and Z boson states.  This is done by employing eigenstates of the covariant lattice Laplacian, in addition to the Higgs and lattice link variables, to construct gauge invariant vector boson creation operators.  Diagonalizing the transfer matrix in the subspace of Hilbert space spanned by this set yields a massless photon and massive Z particle, as expected.  But in the numerical data there is evidence for more vector bosons in the spectrum, albeit with considerable uncertainty in their masses, with the lowest finite mass particle in the range of 3-4 GeV. \end{abstract}

%
%
%
\maketitle
 
\singlespacing

\section{\label{Intro} Introduction}

   Particles in the spectrum of any non-trivial quantum field theory are extended objects which include, depending on the theory,
surrounding color electric and Higgs fields, and a cloud of virtual particles.  In this sense even an ``elementary'' particle in gauge Higgs theories can be regarded as a composite object.  But bound composite objects in quantum mechanics tend to have a discrete spectrum, leading to the speculation that elementary particles may themselves have a spectrum of excited states.  In 
\cite{Greensite:2020lmh} I have presented numerical evidence that the gauge and Higgs fields surrounding a static charge in SU(3) gauge Higgs theory do have such a spectrum of excitations in the Higgs phase.  In the absence of a lattice formulation
of chiral fermions one cannot carry out a similarly reliable calculation in the electroweak sector of the Standard Model.  One may,
however, ask whether there might be new and unanticipated vector boson states in the quenched electroweak theory, aside from the usual photon, W, and Z bosons.\footnote{This question was recently addressed in \cite{Gangwani:2023dye}, in a version with a fixed magnitude Higgs field, and there we encountered difficulties with scaling.  The present article improves on this situation in several ways, using (i) the standard Higgs potential rather than a fixed-modulus Higgs; (ii) a less ambiguous procedure, namely the solution of the generalized energy eigenvalue problem, to determine masses of the low-lying excitations; and (iii) a much larger set of pseudomatter fields, as discussed below.}

    In this investigation, as in \cite{Greensite:2020lmh}, the use of what we have elsewhere called ``pseudomatter'' fields
\cite{Greensite:2017ajx}  is essential. 
A pseudomatter field is a functional of the gauge fields alone which transforms like a field in the fundamental representation of the gauge group,
except that, unlike a true matter field, it is invariant under the global center subgroup of the gauge group.  These fields can be combined with
standard gauge and matter fields to generate physical states which are invariant under infinitesimal gauge transformations, and yet transform
non-trivially under the global center subgroup of the gauge group.  States of this kind are charged states.  Indeed, since an uncharged (= color neutral)
state is gauge invariant, a charged state by contrast must transform in some way under the gauge group, and yet conform to the Gauss law constraint on physical states.  An example appears
in the simple case of a static charged source coupled to the quantized Maxwell field.  The ground state of this system in $A_0=0$ gauge was derived
long ago by Dirac \cite{Dirac:1955uv}.  In an infinite volume this is
\beq
  \Psi_{\text{chrg}} =  \pbar(x) \rho(x;A) \Psi_0  \ ,
 \eeq
 where $\Psi_0$ is the ground state, $\pbar$ creates a static fermion, with
 \bea
       \rho(x;A) &=&  \exp\left[-i {e\over 4\pi} \int d^3z ~ A_i(\vz) {\pa \over \pa z_i}  {1\over |\vx-\vz|}  \right]
 \label{rho}
 \eea
 and \cite{Misner:1973prb}
 \beq
        \Psi_0[A] =  \exp\left[ - \int d^3x \int d^3y ~  
         { \nabla \times {\bf A}(x) \cdot \nabla \times {\bf A}(y) \over 16\pi^3 |x-y|^2} \right] \ .
\label{MTW}
\eeq
The operator $\rho(x;A)$ is an example of a pseudomatter field.  It is easy to check that this field transforms covariantly
except under global U(1) transformations which, of course, do not transform the gauge field.  As a result, under global
U(1) transformations $g(x) = e^{i\th}$ which do not depend on position
\beq
\Psi_{\text{chrg}} \ra e^{-i\th} \Psi_{\text{chrg}}  \ .
\eeq
It is this covariance of a physical state under the global center subgroup of the gauge group which distinguishes, in an infinite volume, charged from neutral states in the massless and confining phases of a gauge Higgs theories.  If the global center subgroup is unbroken, all charged and neutral states are orthogonal.  This distinction breaks down when the global center subgroup of the gauge group, ``GCS'' hereafter,
is spontaneously broken, and in that case the system is in the Higgs phase, as argued in \cite{Greensite:2020nhg}.

    These considerations extend to non-abelian theories, where the gauge transformations in the GCS of the SU(N) gauge group consist of elements of the $Z_N$ center, and we consider gauge Higgs theories with the Higgs field $\phi$ transforming in the fundamental representation of the gauge group.  Pseudomatter operators can be used to construct gauge transformations to a physical gauge, and in fact, in the abelian theory, the pseudomatter field 
$\rho^\dg(x;A)$ transforms the abelian gauge field to Coulomb gauge.   Another example on the lattice is the set of
eigenstates $\xi_n^a(x;U)$ of the covariant lattice Laplacian on a time slice, where
 \beq
            -D^{ab}_{xy}[U] \xi_n^b(y;U) = \l_n \xi_n^a(x;U) \ ,
 \eeq
 and  
 \bea 
    D^{ab}_{\vx \vy} &=& \sum_{k=1}^3 \left[2 \d^{ab} \d_{\vx \vy} - U_k^{ab}(\vx) \d_{\vy,\vx+\hat{k}}  - U_k^{\dg ab}(\vx-\hat{k}) \d_{\vy,\vx-\hat{k}}   \right]  \non \\
\label{Laplacian}
\eea
is the lattice Laplacian (superscripts are color indices).  Note that since the lattice gauge field $U_\m(x)$ is unaffected by the global center subgroup of the gauge group, so are the $\xi_n$.  Using the pseudomatter operators $\xi_n$, or any other pseudomatter operators, we can construct in an infinite volume physical states in gauge Higgs theories such as
\beq
  \Psi_{\text{chrg}} =  \pbar^a(x) \xi^a(x;U) \Psi_0 [U,\phi]  \ ,
\eeq
which transform covariantly under the GCS, and are therefore charged and orthogonal to all neutral states, e.g. 
\beq
          \Psi_{\text{neutral}} = \pbar(x) \phi(x) \Psi_0  \  ,
\eeq
providing this symmetry is not spontaneously broken.  We view the Higgs phase as the phase in which the GCS symmetry 
{\it is} spontaneously broken, as stated above, and the sharp distinction between charged and neutral states no longer exists; for details cf.\ \cite{Greensite:2020nhg}.\footnote{An exception is the existence of electrically charged states in the Higgs phase of the electroweak theory.  We explain how this works out in our framework in an appendix.}  In a finite volume we must consider two operators, transforming in opposite ways under the GCS acting on the vacuum, e.g.
\beq
 \Psi_{xy} = \pbar^a(x) \xi^a(x;U)  \xi^{\dg b}(y;U) \psi^b(y) \Psi_0 [U,\phi]  
\eeq
or more generally
\beq
 \Psi_{xy} = \pbar^a(x) V(x,y;U) \psi^b(y) \Psi_0 [U,\phi]  
\eeq
with $V(x,y,U)$ a functional of the lattice gauge field transforming as  $V(x,y,;U)\ra g(x) V(x,y,U) g^\dg(y)$ under a gauge transformation.
An isolated charge at point $x$ is obtained with increasing volume by taking ${y \ra \infty}$.

   Our strategy is to combine the Higgs and $\xi_n^a(x;U)$ pseudomatter fields with lattice link variables to construct a set of gauge invariant operators, which, operating on the ground state, span a finite subspace of physical states corresponding to neutral vector bosons. We then diagonalize the transfer matrix in this subspace, and study whether the masses of some of the excitations are stable with respect to expanding the dimensionality of the subspace.  It will be seen in section \ref{Results} that at least one new state of this kind appears, in addition to  the photon and Z. 

\section{Procedure}

   In this article we will be concerned only with the bosonic part of the electroweak sector of the Standard model since, as already mentioned, we do not yet have a satisfactory formulation of chiral gauge theories.  Lattice treatments of the bosonic part of the electroweak theory go back to Shrock  \cite{Shrock:1985ur}; see also the more recent work by Veselov and Zubkov  \cite{Zubkov:2008gi}.  The lattice action is
 \bea
 S  &=& -\b \sum_{plaq} \left[\oh \tr[UUU^\dg U^\dg] + {1\over \tan^2(\th_W)} \text{Re}[VVV^\dg V^\dg] \right] \non \\
      & &  -2\sum_{x,\m} \text{Re}[\phi^\dg(x) U_\m(x) V_\m(x) \phi(x+\hat{\m})]\non \\
      & & + \sum_x \{ -(\g-8) \phi^\dg(x)\phi(x)  + \l (\phi^\dg(x)\phi(x))^2 \} \ ,
 \eea
 with SU(2) gauge field $U_\m(x)$, U(1) gauge field ${V_\m(x) = e^{i\th_\m(x)}}$, and Higgs field $\phi(x)$, with $\th_W$ the Weinberg angle.

    For the parameters of the lattice theory, we take the Weinberg angle to be $\sin^2\th_W = 0.231$ \cite{rpp}.  The Weinberg angle is scheme dependent, but the small variation of this parameter with scheme will not concern us here.  Then we have \cite{Melo:2017agn}
\beq
           \b = {4\sin^2 \th_W\over e^2} = 10.1 ~~~,~~~ \l = 0.13
\label{couplings}
\eeq
which completes the definition of the lattice theory we will study.  $\g$ is a free variable on the lattice, which in the continuum action
has dimensions of mass squared.  Different values of $\g$ on the lattice correspond to different lattice spacings.  We will compare results at $\g=2,4,8$.

   Define $\tilde{U}_\m(x) = U_\m(x) V_\m(x)$, with the lattice Laplacian operator $D^{ab}_{xy}[\tilde{U}]$ covariant under the SU(2)$\times$U(1) group.    Let $\zeta_{1}(x)=\phi(x)$ be the Higgs field, and ${\zeta_i(x)=\xi_{i-1}(x), i=2,3,...,n_{ev}+1}$ 
be the lowest $n_{ev}$ eigenstates of the SU(2)$\times$U(1) covariant Laplacian operator.  Define
\bea
    \eta(\vx) e^{i{\cal A}^i_\m(\vx)} = \zeta_i^\dg(x) \tilde{U}_\m(\vx,t) \zeta_i(\vx+\hat{\m})  \ ,
\eea
with $\eta(\vx)>0$ and
\beq
     A^i_\m(\vx) = \sin({\cal A}^i_\m(\vx)) \ .
\eeq
We identify $Z_\m(x) \equiv A^1_\m(x)$ as the operator which creates Z bosons \cite{Zubkov:2008gi}.  Next,  
\bea
     Q_\m^i &=& {1\over L^3} \sum_\vx A^i_\m(\vx) \non \\
     |\Phi_\m^i\rangle &=& Q^i_\m |\Psi_0\rangle \ ,
\label{Q}
\eea
where index $\m=1,2,3$ are spatial directions, and $i$ labels the choice of pseudomatter (or Higgs) field 
$\zeta_i$.   

    Now we have an $n_{ev}+1$ dimensional subspace of the full Hilbert space spanned by the $\{|\Phi_\m^i\rangle\}$.  The next step is
to diagonalize the transfer matrix in this subspace, and extract the energy spectrum. Let $T=e^{-Ha}$  be the transfer matrix, $\E$ 
the vacuum energy, and let
\beq
            \t = T e^{\E a} = e^{-(H-\E)a} \ .
\eeq
In general, for any two operators $A,B$ and states
$|A\rangle = A|\Psi_0\rangle$ and $|B\rangle = B|\Psi_0\rangle$, we have 
\beq
           \langle A| \t^t |B \rangle \equiv \langle A| e^{-(H-\E)t}|B \rangle = \langle A^\dg(t) B(0) \rangle \ .
\eeq
 
 Denote by $\Psi_\m^n$ the states which diagonalize $\t$ in the $n_{ev}+1$ dimensional subspace.  These are obtained by first computing, via lattice Monte Carlo simulations, the ${(n_{ev}+1)\times (n_{ev}+1)}$ matrices
\bea
            O_{ab} &=&  \langle\Phi_\m^a|\Phi_\m^b\rangle = \langle Q^{a\dg}_\m(t) Q^b_\m(t) \rangle \non \\
            T_{ab}  &=& \langle \Phi_\m^a|\t|\Phi_\m^b\rangle =  \langle Q^{a\dg}_\m(t+1)  Q^b_\m(t) \rangle \ .
 \eea
 Then we solve numerically the generalized eigenvalue equation
 \beq
              T_{ab} v^n_b = \l_n O_{ab} v^n_b \ ,
 \eeq
 or
 \beq
            [T] \vec{v}^n = \l_n [O] \vec{v}^n \ .
 \eeq
 There will be $N+1$ vectors $\vec{v}^n$ which satisfy this equation, and then
 \beq
              |\Psi^n_\m\rangle = \sum_a v^n_a |\Phi_\m^a\rangle 
 \eeq
 are the eigenstates of $\t$ in the subspace.   These states are zero momentum by construction.  If $|\Psi^n\rangle$ is a one particle
 state, then the corresponding mass is $M_n = -\log(\l_n)$.  We will refer to this generalized eigenvalue problem calculation as
 the GEP procedure.

    Of course, the $\l_n$ arrived at in this way are the eigenvalues of the transfer matrix in a finite subspace of Hilbert space, and obviously
may differ markedly from the eigenvalues in the full Hilbert space.\footnote{The standard method to arrive at better estimates of the spectrum in the full Hilbert space, via the generalized eigenvalue problem, involves computing correlators $\langle Q^{a\dg}_\m(t+t_0) Q^b_\m(t_0) \rangle$  at long time extensions $t$, as
explained in \cite{Luscher:1990ck,Blossier:2009kd}.  Unfortunately we find that in our particular case, with the data that has been collected, this method is overwhelmed by statistical error.}  We observe, however, that in principle the exact spectrum
is obtained by this method in the limit that the $\{|\Phi_n\rangle \}$ spans the full Hilbert space, and so we expect that the low-lying eigenvalues of the transfer matrix in the finite subspace will tend towards the corresponding eigenvalues in the full Hilbert space as the dimensionality of the subspace increases.  So in our approach what we look for is the convergence of the low-lying mass spectrum with increasing $n_{ev}$.

\section{Results \label{Results}}

\subsubsection{Simulations}

The lattice Monte Carlo simulations were run on lattice volume $16^3 \times 72$, and also a smaller volume $12^3 \times 72$,  at the couplings
shown in \rf{couplings}, and $\g=2,4,8$.  100,000 thermalizing sweeps were followed by 400,000 sweeps, with data taken every
200 sweeps.  Error bars were obtained by running ten independent simulations at each $\g$ and lattice volume.  

\subsubsection{Transition to the Higgs phase}

\begin{figure}[htb]
 \includegraphics[scale=0.6]{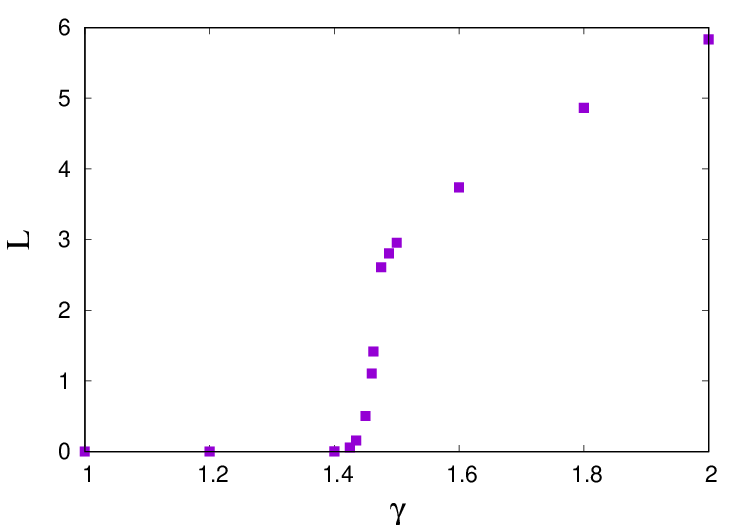}
 \caption{Expectation value of the gauge invariant link $L$ defined in eq.\ \rf{L} vs.\ $\g$ on a $12^4$ lattice volume.}
 \label{zphase}
\end{figure}

     First we must be sure to work in a range of $\g$ which is in the Higgs, rather than the confinement phase.  The Higgs phase, as we have argued at length elsewhere \cite{Greensite:2020nhg}, is the phase in which the global center symmetry of the group is spontaneously broken; the
transition line may or may not coincide, partially, with a line of thermodynamic transition.  In fact, with the choice of $\b, \l$ and Weinberg angle specified above, one finds, from inspection of the data for 
\beq
   L = \langle \phi^\dg(x) \tilde{U}_\m(x) \phi(x+\hat{\m})\rangle ~~\text{vs.} ~~\g  ~~~,
\label{L}
\eeq
that there is a transition to the Higgs phase, most likely first order, at 
$\g\approx 1.45$.  The data obtained on a $12^4$ lattice is displayed in Fig.\ \ref{zphase}.  Our investigation was carried out well above this transition, at
$\g=2,4,8$.   

Lattice Monte Carlo simulations were used to obtain the correlation functions
\beq
   \langle \Phi_\m^a | \t^t | \Phi_\m^b \rangle = \langle Q^{a\dg}_\m(t_0+t)  Q^b_\m(t_0) \rangle
\eeq
needed to compute the matrices $O_{ab}$ ($t=0$) and $T_{ab}$ ($t=1$), and from there the mass spectrum.  

\subsubsection{The Z boson}

\begin{figure}[htb]
 \includegraphics[scale=0.6]{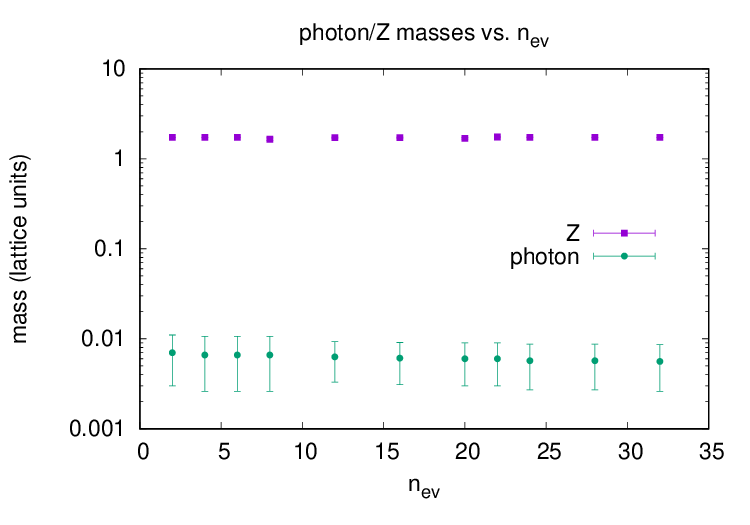}
 \caption{Photon and Z mass in lattice units vs.\ $n_{ev}$ at $\g=4$ on a $16^3\times 72$ lattice volume.}
 \label{pZ}
\end{figure}

   The state which we will identify as the Z boson appears already at $n_{ev}=0$.  At $\g=4$ its mass $m_Z$ in lattice units is $1.735(2)$, and a state of this mass can be identified among the excited levels as $n_{ev}$ increases, as shown in Fig.\ \ref{pZ} for $\g=4$.  For comparison, the tree-level perturbative value on the lattice is
\beq
   m^{tree}_Z =   {1\over \cos \th_W} \sqrt{\g \over \lambda \b} 
\label{mZ}
\eeq
and at $\g=4$ the ratio $m_Z/m^{tree}_Z$ is 0.87, so these values are reasonably close.  We therefore tentatively identify states of this mass, appearing already at $n_{ev}=0$, as the Z boson.

\subsubsection{The photon}

\begin{figure}[htb]
 \includegraphics[scale=0.6]{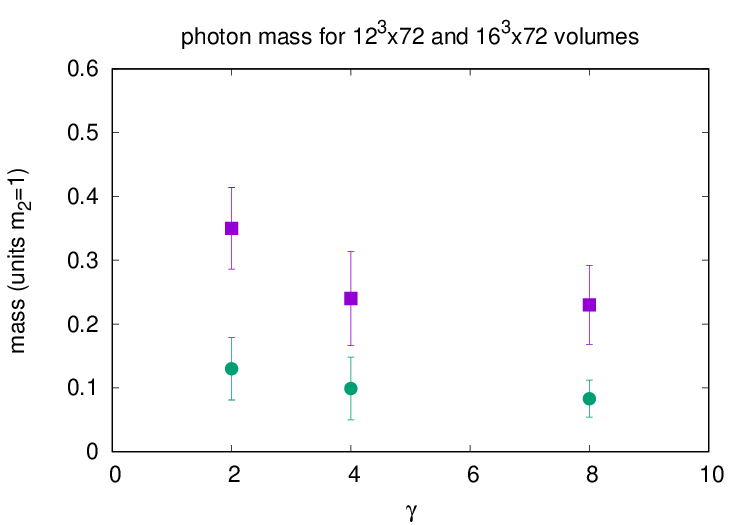}
 \caption{Photon mass vs.\ $\g$ on $12^3\times 72$ and $16^3\times 72$ lattice volumes, at $n_{ev}=32$.  Masses are in units of the
 mass $m_2$ of the next higher excitation.} 
 \label{photon}
\end{figure}

\begin{figure}[htb]
 \includegraphics[scale=0.6]{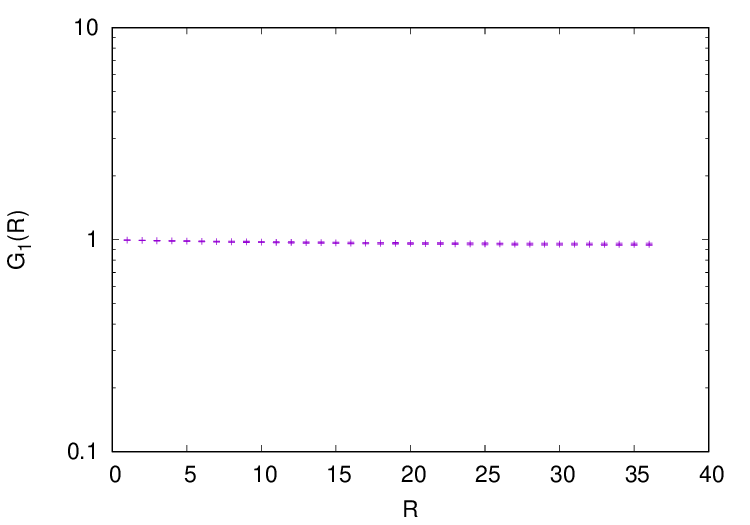}
 \caption{$n=1$ level (photon) Euclidean time correlation function $G_1(T)$ vs.\ time $T$.} 
 \label{G1}
\end{figure}

   At $n_{ev}=2$ a very light mass state appears, and this lowest mass state remains very light as $n_{ev}$ increases, as also shown in Fig.\ \ref{pZ}.   In order to compare masses of the lightest state at different $\g$, let us provisionally choose units in which the next-to-lowest energy state has the value unity, and scale all other energy levels accordingly. We then compare, in Fig.\ \ref{photon}, the mass of this lowest mass state (equivalently the ratio $m_1/m_2$ of the two lowest excitations) on $12^3 \times 72$ and $16^3\times 72$ lattice volumes at $\g=2,4,8$, and we see that the mass drops by about a factor of three at the larger lattice volume. Therefore there is reason to believe that this finite mass for lowest state is just a finite volume effect, and that the lowest state actually represents a massless photon. The very low mass arrived at by the GEP is consistent with a plot of $G_1(T)$ for this lowest state (Fig.\ \ref{G1} at $n_{ev}=32$), where
 \bea
             G_n(t) &=& \langle \Psi_\m^n| \t^t |\Psi_\m^n\rangle \non \\
                        &=& \sum v^{n*}_a v^n_b \langle \Phi_\m^a | \t^t | \Phi_\m^b \rangle  \ ,
 \label{G}
 \eea   
This time correlator is very nearly constant at large $T$ for ${n=1}$, consistent with a near massless particle state.  

\subsubsection{The lightest massive state}

\begin{figure}[htb]
 \includegraphics[scale=0.6]{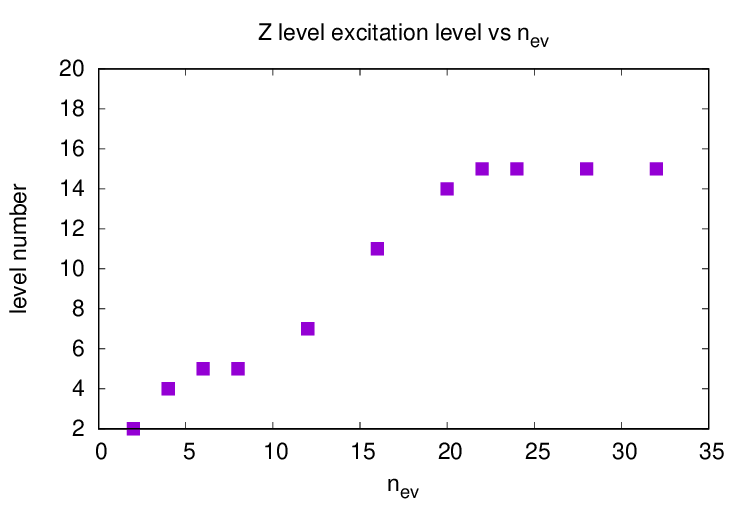}
 \caption{Excitation number of the Z boson vs.\ $n_{ev}$, showing how new excitations appear in the spectrum between
 the photon and the $Z$ with increasing the number of pseudomatter fields.} 
 \label{Zlevel}
\end{figure}

   As $n_{ev}$ increases, more states appear in the spectrum between the photon and the state we identify, by its mass in lattice units, as the Z.  This increase is displayed
in Fig.\ \ref{Zlevel}, where we plot the excitation number of the state identified as the Z versus $n_{ev}$ ($\g=4$, and $16^3\times 72$ volume).  There is a steady increase in the excitation number, stabilizing at $n_{ev}=22$.  As $n_{ev}$ increases further, the additional states all appear at masses higher than the $Z$.  At $n_{ev}=22$ and above, the Z boson is the state at level 15.
 
 \begin{figure}[htb]
 \includegraphics[scale=0.6]{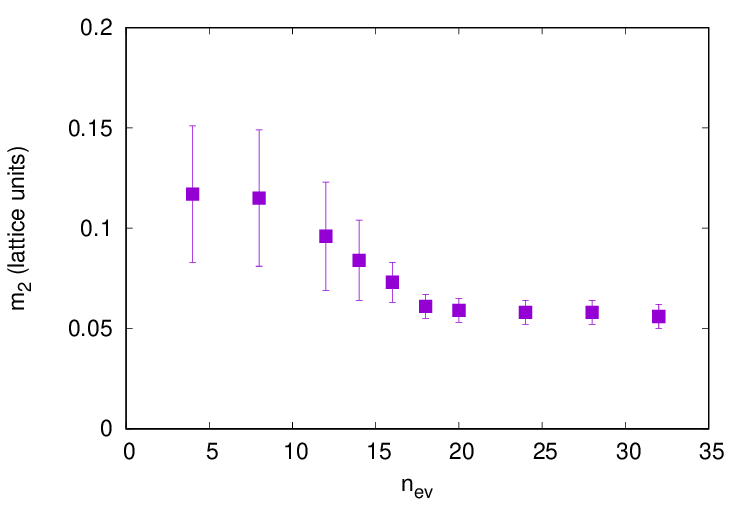}
 \caption{Mass (in lattice units) of the first excitation above the photon state, vs.\ the number $n_{ev}$ of pseudomatter states spanning the truncated
 space of states.  Data is for $\g=4$, and we have convergence by $n_{ev}=20$.} 
 \label{m2}
\end{figure}  

\begin{figure}[htb]
 \includegraphics[scale=0.6]{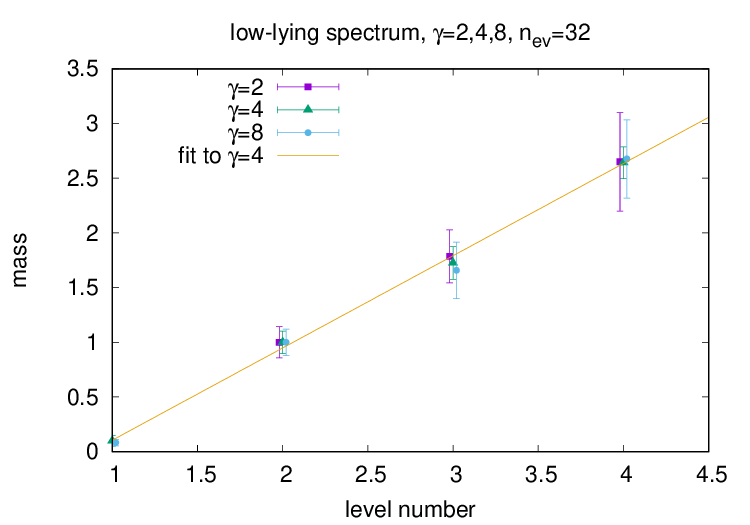}
 \caption{Energies of the first four energy levels in the spectrum obtained from the generalized eigenvalue calculation (with $n_{ev}=32$).
 Results are shown for $\g=2,4,8$, in units where the mass of the second level is $m_2=1$.  The lattice volume is $16^3\times 72$, and
 the straight line is a fit to the $\g=4$ data points. Data points are slightly displaced horizontally for visibility.}
 \label{gall_4}
\end{figure}

    Figure \ref{m2} displays the mass of the first excitation above the photon state in lattice units vs.\ $n_{ev}$, which converges at $n_{ev}=20$.  In Fig.\ \ref{gall_4} we show the first four energy levels in the spectrum at $\g=2,4,8$, in units of the mass of the $n=2$ energy level $m_2$.  This is the level just above the photon state, and we set the scale so that $m_2=1$.  Note that from here on all computations are on a $16^3\times 72$ volume unless otherwise stated.  It should be noted that these first four excitations are fit fairly accurately by a straight line fit to the $\g=4$ data.  Suppose that this line had slope=1 and an $x$-intercept at $n=1$.  Then the interpretation is obvious: the first level is a massless state (the photon), the second is a finite mass particle state of mass $m_2$, the third and fourth levels correspond respectively to two and three particles of mass $m_2$, all
at momentum $p=0$.  However, if these particles have an attractive interaction, then the slope need not be exactly one, we would only expect that slope in the infinite volume limit, where the interaction can presumably be neglected.  In fact the slope of the fitting line on the $12^3 \times 72$ lattice is 0.75(1), while the slope on the $16^3\times 72$ lattice is 0.84(2), closer to the ideal slope=1 value.   Thus we attribute the deviation from slope=1 to a finite size effect due to particle interactions, and interpret the third and fourth levels as two and three zero-momentum particle states, respectively, each of mass
$m_2$.    
    

   Now the question is what is actually the mass of the lowest excitation above the photon, at level $n=2$.  If we look only at the $\g=4$ data, and taking
the result seen in Fig.\ \ref{pZ} that $m_Z \approx 1.74$ in lattice units, and given the physical mass of the Z boson at 91.2 GeV, then the mass of the level 2  excitation comes out to  $m_2=3.0(3)$ GeV.  This cannot be identified as two photons of opposite momenta, because for a lattice of spatial extension 16 this would be an energy of at least ${4\pi\over 16} = 0.785$ in lattice units.   For comparison, the mass of $m_2$ in lattice units at $\g=4$ is 0.056, and the ``two photon'' interpretation is untenable.  We must be looking at a single particle state.

\subsubsection{Uncertainties}

\begin{figure}[htb]
 \includegraphics[scale=0.6]{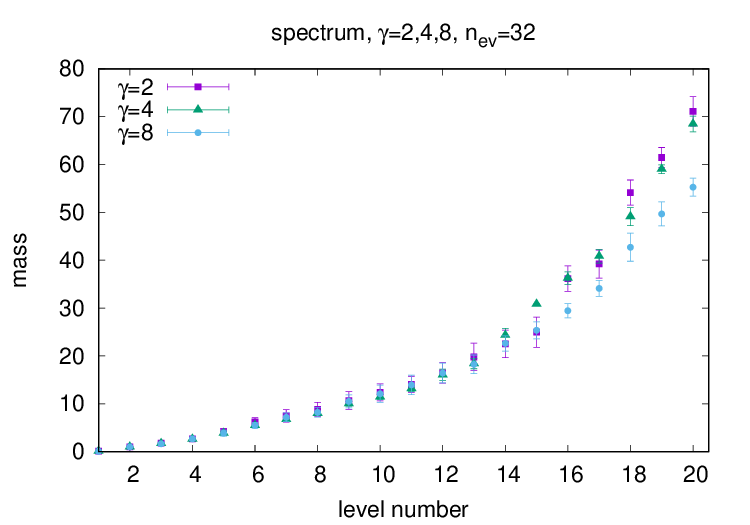}
 \caption{Same as Fig.\ \ref{gall_4}, this time displaying 20 energy levels.  Note that the close agreement among levels found for different $\g$ at
 the lower levels in the spectrum is not maintained at the upper end of the spectrum.}
 \label{gall_20}
\end{figure}

   Unfortunately the results at other $\gamma$ values and $n_{ev}=32$ complicate the picture.  In Fig.\ \ref{gall_20} we show the spectrum for $\g=2,4,8$ up to level 20, again with the second level excitation normalized to unity.  It is clear that the spectra do not agree at the higher excitation levels, in particular at level $n=15$ which should be the Z boson, and this calls into question the use of $\g=4$ to set the physical scale.  If we use also the level 15 masses at $\g=2$ or $\g=8$ to set the physical scale, then we get a different result.  In fact we find in all, also including the same analysis on data from the smaller $12^3\times 72$ volume, the results for the mass $m_2$ shown in Table I. \bigskip

\begin{table}[t!]
\begin{center}
\begin{tabular}{|c|c|c|} \hline
         $m_2$ (GeV) &  $ \g $ & lattice volume  \\
\hline
        3.6(5) &          2  &  $16^3 \times 72$  \\
        3.0(3) &          4  &  $16^3 \times 72$  \\
        3.6(4) &          8  &  $16^3 \times 72$ \\
        4.0(4) &          2  &  $12^3 \times 72 $ \\
        3.5(3) &          4  &  $12^3 \times 72$  \\
        3.3(3) &          8  &  $12^3 \times 72$  \\       
\hline
\end{tabular}
\caption{Mass $m_2$ of the lightest state above the photon, computed at three different $\g$ values and two
lattice volumes.  In each case the physical scale is set by identifying the level 15 state with the Z boson (see text).} 
\label{tab1}
\end{center}
\end{table}
 The most that can be said from this data is that we appear to have a massive excitation above the photon with a mass somewhere between 3 and 4 GeV.
Obviously this is a substantial uncertainty in mass, reflecting the mentioned disagreement seen in Fig.\ \ref{gall_20} at the higher end of the spectrum.
 
\section{Conclusions}

    We have diagonalized the transfer matrix in the quenched electroweak theory in a finite dimensional subspace of physical states, and found a spectrum of vector boson states which appears to include both the photon and Z boson. The subspace is obtained using gauge invariant operators, constructed from lattice link variables combined with either the Higgs field or pseudomatter fields, operating on the vacuum. One might expect that the only other states in the spectrum would be multiparticle states consisting of photons and Z bosons, with finite lattice momenta summing to zero.  Instead we find a spectrum of states between the massless photon and Z boson which cannot be interpreted as multi-photon states, and the lightest particle state above the photon is surprisingly light compared to the Z boson.  Our estimate of the mass of this particle is subject to a substantial uncertainty, but all indications are that this mass is in the range of three to four GeV.  The origin of such a small mass scale in the spectrum is unclear.

    It is probably premature at this stage, with significant uncertainties in mass and ignorance of width, to search the particle tables for evidence
of a peak corresponding to such a state.  The methodology also comes with caveats.  First, the spectrum of the transfer matrix is computed in a subspace of vector meson states, and in a subspace of finite dimension this may differ significantly from the spectrum in the full Hilbert space.  We have used convergence of the spectrum with increasing dimensionality of the subspace as a criterion that the dimensionality of the subspace is large enough, but of course this could be misleading.  Secondly, it must be remembered that the lattice theory itself, being ultimately a $\phi^4$ theory, does not have a non-trivial continuum limit, and reducing the $\gamma$ parameter while keeping other couplings fixed ultimately runs into a transition to the confined phase.
Finally, the computation is carried out in the quenched electroweak theory, and this will remain a limitation, in comparing to experiment, until the problem of formulating lattice chiral fermions is resolved.

    What can be done at present is to reduce uncertainties in the existing GEP calculation, and ensure the robustness of the result.  We have used lattice volumes of $12^3\times 72$ and $16^3\times 72$ and found reasonably consistent results, but computations on larger spatial volumes, more $\g$ values, and higher dimensional subspaces with $n_{ev}$ substantially greater than 32 are desirable.  It may also be helpful to use smeared link variables in the construction.  In particular we would like to see if the calculation can be improved to the point that the GEP results at different $\g$ values will come into closer agreement at the higher end of the spectrum (perhaps requiring $\g$-dependent renormalization of the lattice-scale couplings), since this is the main source of uncertainty in the mass of the lightest state.  That effort will be the focus of future work.

 \acknowledgments{I thank George Fleming for a helpful discussion.  This research is supported by the U.S.\ Department of Energy under Grant No.\ DE-SC0013682.}   
 
 \appendix*
 \section{Charged states in the electroweak theory}
 
In the electroweak gauge Higgs theory, the global center subgroup $Z_2 \times$U(1) is spontaneously broken in the Higgs phase, as detected by
the gauge-invariant order parameter for GCS breaking described in ref.\ \cite{Greensite:2020nhg}.  This raises the question, in view of the discussion in the Introduction, of how there can be electrically charged states in the Higgs phase, orthogonal to all
neutral states.

The answer is that in the Higgs phase of the electroweak theory there is still an unbroken global symmetry, and one can construct physical states which satisfy the Gauss Law constraint and also transform covariantly under this symmetry. But in the electroweak theory the symmetry is not exactly the GCS. 
The key (and common) observation is that in unitary gauge there is a remnant U(1) local gauge symmetry, consisting of gauge transformations
\beq
           g(x; \th(x)) = e^{i\th(x)/2} e^{i\th(x) \s_3/2} \ ,
\label{remnant}
\eeq
and that this symmetry is unbroken in the Higgs phase.  Since this is still a local symmetry, which in view of the Elitzur theorem is unbreakable, it is 
is only a global subgroup of this remnant symmetry, consisting of transformations
\beq
         g(x,\th) = g(\th) =  e^{i\th/2} e^{i\th \s_3/2} 
\eeq
which could in principle, but does not, break spontaneously.  If $G[x;\phi(x)]$ is the gauge transformation taking any configuration into unitary gauge, then we can also write this global symmetry, acting on any configuration (not necessarily in unitary gauge), as
\beq
        g(x,\th) = e^{i\th/2} e^{i\th \s_3/2} G[x,\phi(x)] \ .
\label{global}
\eeq

    Now suppose $\psi$ is a matter field which, like the left-handed fermions in the Standard Model, is in the same SU(2) group representation as the Higgs field, but with opposite hypercharge.  Then 
\beq
          \Psi_{neutral} = \pbar(x) \s_2 \phi^\dg(x) \Psi_0 \ ,
\eeq
where $\Psi_0$ is the vacuum state and $\s_2$ is a Pauli matrix, is an electrically neutral state invariant under all gauge transformations including \rf{global}.   On the other hand, if $A_\m$ is the U(1) gauge field of the SU(2)$\times$U(1) gauge Higgs theory, and $\rho(x;A)$ is the pseudomatter field shown in \rf{rho}, let
\beq
      \Psi_{chrg} = \pbar(x) \phi(x) \rho(x,A) \Psi_0 \ .
\eeq
This is a physical state, invariant under all infinitesimal gauge transformations, which nonetheless transforms covariantly under the global subgroup
\rf{global}, with the state transforming as
\beq
      \Psi_{chrg} \ra e^{-i\th} \Psi_{chrg} \ ,
\eeq
and is therefore orthogonal to all electrically neutral states.  A similar construction can be carried out for electrically charged $W$ particles.  So
there is still a sharp distinction in the electroweak theory between electrically
charged and uncharged states in the Higgs phase, quite similar to the situation for the abelian theory described in the Introduction. 

 It should be noted that the idea of constructing gauge invariant particle states in the electroweak theory, and electrically charged states which transform only under a U(1) symmetry, goes back to the old work in refs.\ \cite{Frohlich:1981yi,tHooft:1979yoe,Banks:1979fi}, and \cite{Frohlich:1981yi} in fact alludes to the necessity of attaching string operators to create physical electrically charged particle states.  In this article that suggestion is made concrete,
but dropping string operators (presumably Wilson lines) in favor of the pseudomatter operators described above.

 \bibliography{sym3}

\end{document}